# Pressure induced metallization and loss of surface magnetism in FeSi

Yuhang Deng[1], Farhad Taraporevala[1], Haozhe Wang[2], Eric Lee-Wong[3], Camilla M. Moir[1], Jinhyuk Lim[4], Shubham Sinha[4], Weiwei Xie[2], James Hamlin[4], Yogesh Vohra[5], and M. Brian Maple[1]

[1]*Department of Physics, University of California, San Diego, La Jolla, California 92903, USA*

[2]*Department of Chemistry, Michigan State University, East Lansing, Michigan 48824, USA*

[3]*Department of NanoEngineering, University of California, San Diego, CA 92093, USA*

[4]*Department of Physics, University of Florida, Gainesville, FL 32611, USA*

[5]*Department of Physics, University of Alabama at Birmingham, Birmingham, AL 35294, USA*

*Corresponding author: M. Brian Maple. Email: mbmaple@ucsd.edu

**Abstract**

Single crystalline FeSi samples with a conducting surface state (CSS) were studied under high pressure ($P$) and magnetic field ($B$) by means of electrical resistance ($R$) measurements to explore how the bulk semiconducting state and the surface state are tuned by the application of pressure. We found that the energy gap ($\Delta$) associated with the semiconducting bulk phase begins to close abruptly at a critical pressure ($P_{cr}$) of ~10 GPa and the bulk material becomes metallic with no obvious sign of any emergent phases or non-Fermi liquid behavior in $R(T)$ in the neighborhood of $P_{cr}$ above 3 K. Moreover, the metallic phase appears to remain at near-ambient pressure upon release of the pressure. Interestingly, the hysteresis in the $R(B)$ curve associated with the magnetically ordered CSS decreases with pressure and vanishes at $P_{cr}$, while the slope of the $R(B)$ curve, $dR/dB$, which has a negative value for $P < P_{cr}$, decreases in magnitude with $P$ and changes sign at $P_{cr}$. Thus, the CSS and the corresponding two-dimensional magnetic order collapse at $P_{cr}$ where the energy gap $\Delta$ of the bulk material starts to close abruptly, revealing the connection between the CSS and the semiconducting bulk state in FeSi.

**Introduction**

Kondo insulators (KIs) comprise a class of correlated *f*-electron lanthanide- and actinide-based compounds that have challenged and fascinated researchers for decades. Much of the interest in these materials is driven by two striking features of the ground state – it is (1) insulating with a small energy gap $\Delta \sim 10$ meV and (2) nonmagnetic, despite emerging from a dense Kondo lattice of localized *f*-electron magnetic moments. It is believed that the localized *f*-electrons hybridize with conduction electrons to form a hybridization band with a small excitation gap (referred to as a hybridization gap or Kondo gap) within which the Fermi level is located [1]. The *f*-electron magnetic moments are completely screened and cannot interact to form a magnetically ordered state via the Ruderman-Kittel-Kasuya-Yosida (RKKY) interaction that is responsible for magnetic order in lanthanides and uranium intermetallics. Recent years have seen a resurgence of research on KIs due to the discovery of certain topological insulators (TIs) [2, 3, 4] and thus the possibility of finding topological Kondo insulators (TKIs) – a strongly correlated electron version of a normal TI [1, 5]. The nature of KIs such as large spin-orbit coupling of *f*-electrons and the opposite parities of the *f*-states (odd parity) and the *d*-states (even parity) that contribute to the conduction band make KIs promising candidates for $Z_2$ TIs [5]. Pressure is often used to tune the physical properties of KIs and TIs [6, 7, 8]. For

KIs, pressure can affect both the strength of the hybridization $V$ which is responsible for the opening of the Kondo gap and the bandwidth $W$ of the hybridization bands [5, 9], producing complicated variations of the gap under pressure and eventually the closure of the gap at sufficiently high pressure which is sometimes accompanied by the emergence of magnetic order, as in the case of SmS and $SmB_6$ [10, 11, 12].

The correlated electron small gap semiconductor FeSi has attracted a great deal of interest since it was first studied at Bell Laboratories in the 1960s [13]. The compound undergoes continuous metal-semiconductor [14] and magnetic-nonmagnetic [13, 15, 16] transitions with decreasing temperature $T$. The behavior of the transport, thermal, and magnetic properties and spectroscopic measurements revealed the opening of an energy gap $\Delta$ with decreasing $T$ [17] that led to the conjecture that FeSi may be a $d$-electron counterpart of an $f$-electron KI [14, 18]. In 2018, our group reported the observation of a peak in the electrical resistance $R(T)$ of FeSi flux grown single crystals at $T_s = 19$ K, followed by metallic behavior in $R(T)$ at lower temperatures $T < T_s$, which was attributed to a conducting surface state (CSS), suggesting the possibility that the CSS was of topological origin and that FeSi is a $d$-electron version of an $f$-electron TKI [1, 5, 19, 20]. The existence of a CSS on FeSi was also confirmed in a tunneling spectroscopic study [21] and an electrical transport experiment using a Corbino disk geometry [22]. We also found that the behavior of the magnetoresistance of FeSi is similar to that of $SmB_6$, which is regarded as a prototypical TKI [20]. Recently, we reported magnetoresistance ($MR$) measurements on FeSi that revealed asymmetry and hysteresis in $R(B)$ data in the CSS temperature region $T < T_s$ [23]. Measurements of the angular dependence of the magnetoresistance showed anisotropy with two-fold rotational symmetry in the temperature region of the CSS and isotropic behavior at higher temperatures where the resistivity is dominated by the semiconducting bulk state [23]. This behavior suggested that the CSS of FeSi may actually support two-dimensional (2D) ferromagnetic (FM) order [23], as found in another TKI candidate $SmB_6$ [24]. A very recent paper by Avers *et al.* confirmed the surface ferromagnetism in Te flux grown FeSi single crystals by using Corbino disk electrical transport methods and magnetization measurements on size selected FeSi fragments [25].

In this paper, we report the results of an investigation of the electrical resistance $R$ of FeSi as a function of magnetic field $B$ and pressure $P$ in a diamond anvil cell (DAC) to ~20 GPa. The purpose of the experiments was to study the transition from the small gap semiconducting state to the metallic state expected to form at sufficiently high pressure [20]. In particular, we wanted to determine whether the energy gap $\Delta$ in the bulk semiconducting material closes discontinuously or continuously with pressure at the critical pressure $P_{cr}$. We were especially interested in the second case where $P_{cr}$ would be a quantum critical point (QCP) in the vicinity of which emergent phases (e.g., unconventional superconductivity, exotic type of magnetic order) and/or non-Fermi liquid behavior in $R(T)$ (e.g., $R(T) \sim T$, specific heat $C$ divided by $T$, $C(T)/T \sim \ln T$) are often found. We also wanted to see how the CSS and the apparent 2D FM order change at $P_{cr}$. While we succeeded in closing the energy gap with pressure between 10 - 15 GPa, the transition to the metallic state was abrupt and indicative of a discontinuous quantum phase transition rather than a continuous transition with a QCP. A more detailed study of the actual transition from the semiconducting to the metallic state will require more hydrostatic pressure conditions and access to temperatures lower than the 3 K limit in the present experiments.

**Results and Discussion**

A total of four successful runs were conducted including three runs for FeSi single crystals (runs #4, #5, #7) and one run for powdered FeSi (run #6). In this Letter, we show representative data from run #7; data from runs #4, #5, and #6 are included in the Supplemental Material [26]. In Fig. 1, we display the temperature dependent resistance of FeSi at high pressures and zero magnetic field, as plots of $R$ vs $T$ and $\ln R$ vs $1/T$, respectively. At the first six lower pressures, $R(T)$ of FeSi has a typical semiconducting/insulating characteristic until the emergence of a peak at $T_s$ of ~20 K below which the resistance decreases with cooling. This transition reflects a gradual takeover of the electrical transport by the appearance of the CSS, which has been discussed in detail in refs. [19, 20, 23]. At higher pressures $P >$

$P_{cr} = 10.1$ GPa (measured at 3 K), however, the low temperature resistance drops dramatically by two orders of magnitude within a few GPa, accompanied by a broad maximum at $T_n$ between 50 and 150 K. The shape of the broad maximum is qualitatively different than that of the much sharper peak in $R(T)$ at lower pressure, which marks the domination of the CSS. We speculate that $T_n$ signifies the crossover from semiconducting to metallic behavior of bulk state. On the other hand, the $R(T)$ curves at higher pressure more closely resemble those expected for a normal metal – the resistance becomes smaller at lower $T$ – except for a shallow minimum in $R(T)$ at $T_m$ that emerges at $T < 40$ K. The overall transformation of $R(T)$ under high pressures reflects the pressure-induced metallization of the FeSi single crystal. Note that upon release of pressure to 2.6 GPa (measured at 3 K), the $R(T)$ curve indicates that the FeSi sample remains in its metallic bulk state. No change in $R(T)$ was observed when it was remeasured after leaving the sample at 2.6 GPa for two weeks. This result shows that there is a large amount of hysteresis in the pressure-induced metallization of FeSi.

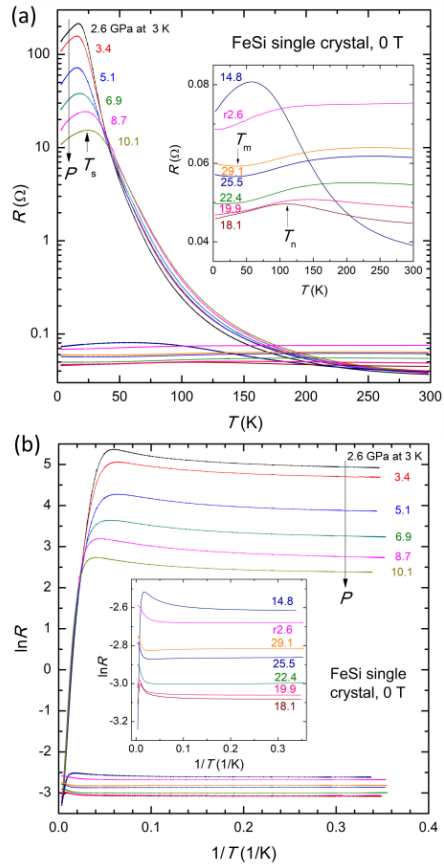

Figure 1. Electrical resistance $R(T)$ data taken on an FeSi single crystal at various pressures between ~3 K and room temperature in zero magnetic field. (a) $R$ vs. $T$ plots at various pressures between 2.6 and 29.1 GPa. Inset: $R$ vs. $T$ plots at 14.8 GPa and higher pressures on an expanded resistance scale for FeSi in the metallic state. The $R$ vs. $T$ curve designated by "r2.6" was taken at 2.6 GPa during release of pressure. (b) ln$R$ vs. $1/T$ plots at the same pressures as in Fig. 1(a). Inset: ln$R$ vs. $1/T$ plots at the same pressures as in the inset of Fig. 1(a) for FeSi in the metallic state. The linear regions at small $1/T$ (large $T$) in the ln$R(1/T)$ curves for FeSi at lower pressures suggest activated behavior, which is described by the Arrhenius function that dominates the electrical transport at higher temperatures. All pressures were measured *in situ* at low temperatures.

Shown in Fig. 2 are magnetoresistance $R(B)$ curves for the FeSi single crystal at different pressures at 3.3 K. Additional data taken at temperatures other than 3.3 K are shown in Fig. S3 in the Supplemental Material [26]. The magnetic field was applied in a fixed direction perpendicular to the direction of the applied ac current and the diamond anvil culet plane. Fig. 2(a) shows the $MR$ in a magnetic field swept between -2 T and 2 T. An obvious loop is seen within ± 0.5 T for pressures below 10.1 GPa, consistent with what was observed for FeSi at ambient pressure [23], which was attributed to a magnetically ordered surface state. The span of the $MR$ hysteresis agrees quite well with that observed in Te flux grown FeSi with the contribution from the bulk state subtracted by using a Corbino disk configuration [25]. With higher pressures, this hysteretic, asymmetrical $MR$ gradually subsides and becomes reversible and symmetrical with respect to the sign of the applied magnetic field. Fig. 2(b) shows the resistance as the magnetic field is ramped monotonically to 9 T. At 2.6 and 3.4 GPa, FeSi has a negative $MR$ in contrast to the $MR$ of a nonmagnetic simple metal, which is positive at low fields. At intermediate pressures of 6.9 and 8.7 GPa, the resistance first decreases and then increases with magnetic field. When $P$ is high enough to metalize FeSi as shown in Fig. 1, the $MR$ finally becomes positive, in accordance with normal metallic behavior. What has been discovered here corroborates the interpretation that the CSS and the associated two-dimensional magnetically ordered state in FeSi depend on the existence of the semiconducting bulk state. Once the semiconducting bulk state is destroyed by the application of pressure, the CSS and the two-dimensional magnetic order are suppressed and simultaneously annihilated (compare Fig. 2(c) and Fig. 3).

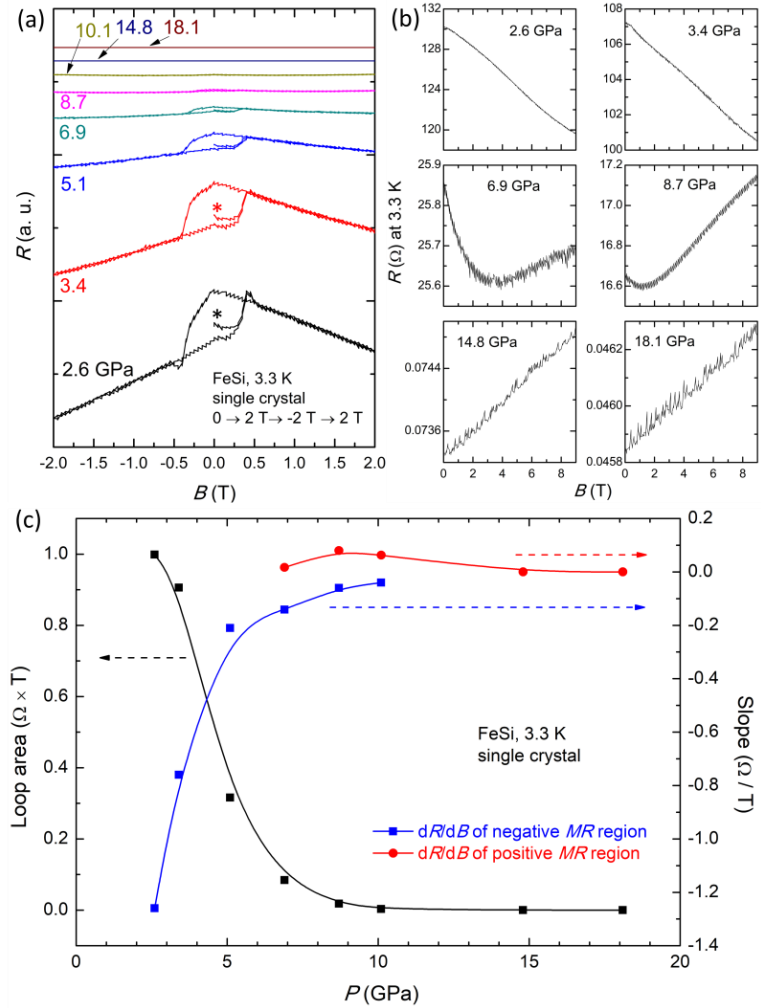

Figure 2. Magnetoresistance $R(B)$ of the FeSi single crystal at various pressures between 2.6 and 18.1 GPa at 3.3 K. (a) $R(B)$ curves taken with the magnetic field $B$ applied in the sequence 0 → 2 T → -2 T → 2 T (an asteroid symbol marking the starting point). Hysteresis in the $R(B)$ curves for FeSi is observed at pressures up to 8.7 GPa, indicating the existence of a magnetically ordered surface state that can be suppressed by pressure. Curves have been shifted vertically for visual clarity. (b) $R(B)$ curves in magnetic fields ramped straight up to 9 T. With increasing pressure, the negative $MR$ gradually evolves into a positive $MR$. The noise in some of the data obtained at higher pressures is due to the change in resistance with magnetic field becoming so small that the measurement noise from the DynaCool ETO module becomes appreciable compared to the change in resistance. (c) $MR$ hysteresis loop area and $dR/dB$ vs. pressure at 3.3 K. Data points are connected by B-spline curves for guides to eye.

Information extracted from the high-pressure transport experiments is summarized in the plots of $R$ vs. $P$ and $T$ vs. $P$ in Fig. 3. The $R$ vs. $P$ data for FeSi at room temperature and 5 K displayed in Fig. 3(a) reveal that $R(298\ \text{K})$ changes very little with increasing $P$, while $R(5\ \text{K})$ undergoes a 2 - 3 orders of magnitude reduction when $P$ increases from ambient pressure to about 15 GPa, above which it remains nearly constant to the highest pressure of ~30 GPa. Interestingly, $R(5\ \text{K})$ does not recover to its original high value when the pressure is released to cross the boundary between the high and low resistance regions marked by the red dashed line in Fig. 3(a). By fitting the two-segment linear region of $\ln R(1/T)$ given in Fig. 1(b) with an Arrhenius function (for more information, refer to Ref. [19]), we obtain the two energy gaps ($\Delta_1$ and $\Delta_2$) for the semiconducting bulk state which are plotted vs. pressure in Fig. 3(b) together with $T_s$, $T_n$ and $T_m$. The energy gap $\Delta_1$ first increases slowly with pressure followed by a rapid decrease. On the other hand, $\Delta_2$ just decreases with pressure before its fitting range becomes too narrow in temperature to calculate its value accurately. This rapid suppression of $\Delta_1$ between 10 and 15 GPa is another feature, in addition to the fundamental change in the shape of $R(T)$, the dramatic suppression of $R(5\ \text{K})$, and the negative-to-positive $MR$ transition, that characterizes the pressure-induced transition to the metallic state of FeSi. Furthermore, we observe the increase of $T_s$ up to ~10 GPa above which the resistance peak is replaced by a broad maximum located at $T_n$ in $R(T)$. The disappearance of $T_s$ represents a competition between the CSS and the pressure induced metallic bulk state – the latter emerging at the expense of the former. From another perspective, this competition provides indirect evidence that the CSS in FeSi is protected by the semiconducting bulk state at low pressures, reminiscent of the topologically protected surface state in the case of a topological insulator. A point worth noting is the pressure evolution of $T_m$ (see resistance minima of $R(T)$ in the inset of Fig. 1(a). This minimum only shows up when there is no peak/broad maximum in the $R(T)$ curves, and like $R(5\ \text{K})$ and $R(T)$ at r2.6 GPa, it maintains its trend even when the pressure has been reduced to below the transition region (10 -15 GPa) between the low-pressure semiconducting bulk state and the high-pressure metallic bulk state.

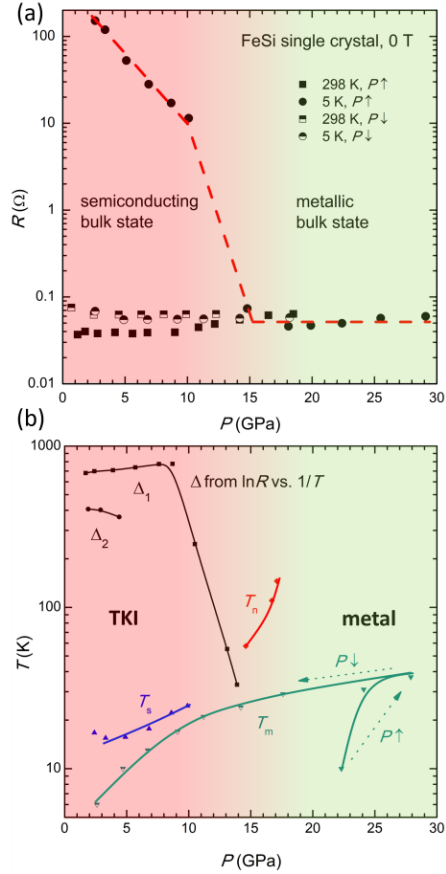

Figure 3. (a) Electrical resistance $R$ vs. pressure $P$ for FeSi at room temperature and 5 K in zero magnetic field. The red dashed lines are guides to eye. (b) Energy gaps derived from the Arrhenius law fits to $\ln R(1/T)$ curves and some characteristic temperatures including $T_s$, $T_n$, and $T_m$, plotted in the $T$-$P$ diagram for the FeSi single crystal. The colored solid lines are guides to eye and divide different states. The dotted arrows point in the direction of applying/releasing pressures. All pressures are measured *in situ* or interpolated. The half-open symbols represent values obtained during releasing pressure. The red and green background colors are used to tentatively separate FeSi into two regions: red for a topological Kondo insulator state and green for a metallic state, with a transition region between 10 and 15 GPa.

Some noteworthy characteristics of the high-pressure metallic state are the following: (1) The value of the electrical resistivity at room temperature is comparable to that of the low-pressure semiconducting phase, which is of the order of $10^{-6}$ $\Omega \cdot m$ [19], much larger than values for a normal metal (~ $10^{-8}$ $\Omega \cdot m$); (2) the electrical resistance $R$ has a very weak temperature dependence and a minimum at low temperature, reminiscent of single-ion Kondo scattering; (3) the magnetoresistance $MR(B, T) \equiv [R(B, T) - R(0\,T, T)]/R(0\,T, T)$ is small ($MR(9\,T, 3.3\,K)$ ~ 2% at 14.8 GPa, and ~ 1% at 18.1 GPa), almost linear, reversible, and positive up to 9 T in contrast to the negative $MR$ in the CSS that is presumably associated with suppression of spin fluctuations by the magnetic field [25, 27]; (4) the semiconducting to metallic state transition appears to be a first-order transition as seen from the discontinuity in $R$ vs. $P$ and $\Delta$ vs. $P$ in Fig. 3. The pressure-induced metallization which occurs above 15 GPa was also observed by Hearne *et al.*, although they measured powdered FeSi with different degrees of disorder and stress introduced by different grinding processes [28]. Besides the similar onset pressure for the metallization, Hearne *et al.* also found that metallic FeSi has a small and positive $MR(8\,T, 2\,K)$ ~ 2% at 15 GPa, and ~ 1.7% at 19 GPa, and that a inflection point or minimum appeared in low temperature $R(T)$ curves when FeSi was entering the metallic state, which was attributed to the onset of ferromagnetic order by analogy with $Fe_{1-x}Co_xSi$ [28].

Despite the similarities mentioned above between our results and those of Hearne *et al.*, there are several notable differences. For example, by studying FeSi powders after different degrees of grinding, Hearne *et al.* discovered that samples with more disorder tend to be metallized at lower pressures [28], whereas our research showed that single crystalline FeSi became metallic at lower pressure than powdered FeSi (compare Figs. 1 and S1 to S2). Moreover, Hearne *et al.* observed a disorder related maximum in $R(T)$ of the metallic FeSi, representing a temperature threshold above which electrons from the valence band edge can be excited to unoccupied states above the Fermi level [28], whereas we did not observe this kind of maximum when FeSi is in the completely metallic state. We stress that the single crystals of FeSi at ambient pressure measured by Hearne *et al.* have a ratio $R(20\ K)/R(300\ K) = 350$ which is much smaller than that of our crystals (4,475), and their FeSi does not show a peak in $R(T)$. Instead, a variable range hopping model was used to describe the low temperature electrical transport behavior of their FeSi samples at ambient pressure [28]. It is interesting that the metallization pressures of the samples studied in these experiments and those of Hearne *et al.* are similar, given the differences in the values of $R(20\ K)/R(300\ K)$ and behavior of $R(T)$ at low temperature observed in the two experiments.

So far pressure induced metallization has been observed in all studied KI's and their hybridization (Kondo) gaps always close with pressure. The compound $SmB_6$ becomes a metal at 4-7 GPa [29, 30, 31] accompanied by magnetic ordering [11]. Additionally, it was proposed that there is an intimate connection between the onset temperature of the CSS of $SmB_6$ and its hybridization gap; in other words, the CSS is protected by the existence of the gap [32]. For $Ce_3Bi_4Pt_3$, pressure suppresses the Kondo insulating gap revealed by gradually flattened $R(T)$ curves under pressure [33], although there seems to be no clear critical pressure for this transition. It was reported that pressure induced metallization in $YbB_{12}$ occurs at ~170 GPa [34]. It is noteworthy that a smaller Kondo gap at ambient pressure does not suggest a lower metallization pressure: $SmB_6$ (40 K and 60 K gaps): 4 GPa [32]; $Ce_3Bi_4Pt_3$ (116 K gap): < 40 GPa [33]; $YbB_{12}$ (28 K and 78 K gaps): 170 GPa [34]. FeSi could be among the first few *d*-electron KIs that have been metallized under pressure. It also serves as a case in which the Kondo gap first increases with pressure, which was also reported in Ref. [35] up to 9.3 GPa. The initial increase of Kondo gaps with pressure was observed in $FeSb_2$ [9], a *d*-electron cousin of FeSi [9, 22], and in $Ce_3Bi_4Pt_3$ which was classified as an electron-type Kondo insulator – a type with the higher-valence *f*-singlet such that pressure favors the *f*-singlet and tends to increase the Kondo interaction and gap [36].

An interesting finding of this study concerns the irreversibility of the high-pressure metallic state which persisted upon release of high pressure on FeSi to near ambient pressure. Our FeSi tin flux grown FeSi single crystals have the expected B20 crystal structure ($\varepsilon$-FeSi), which should not undergo a structural phase transition at room temperature to at least 50 GPa with methanol-ethanol-water or methanol-ethanol as a pressure transmitting medium [37, 38]. At high temperatures, a B20 to B2 (CsCl-FeSi) structural transition has often been observed but there does not seem to be a consistent conclusion about the transition pressure [38, 39, 40]. It is hypothesized that the equilibrium boundary between the B20 and B2 structures could be lower in pressure because a large activation barrier kinetically inhibits the transition [39, 40]. Actually, first-principles pseudopotential calculations show that B2 FeSi is the thermodynamically most stable phase above ~13 GPa [41, 42] and B2 phase FeSi was experimentally observed in thin film form [43, 44, 45] or through high pressure – high temperature synthesis [46]. So, one reasonable explanation for the appearance of the metallic state and its irreversibility is the formation of stress-induced B2 FeSi [47] and its metastability [46] quenched to low pressure – B2 FeSi was calculated to be a metal [47] or a semimetal [42, 44].

Another intriguing possibility is that the sample remained in the B20 phase up to 30 GPa and the semiconductor-metal transition is solely an electronic one but with a very large pressure hysteresis. It is possible because a recent high pressure XRD study revealed the absence of the B2 phase below 36 GPa at room temperature in an extremely nonhydrostatic environment where no pressure medium at all was used for the experiment [47]. Pressure-induced metallization in many materials is accompanied with some crystal structural changes, such as structural phase transitions ($VO_2$ [48]), isostructural phase transitions (SmS [49, 50, 51]), amorphization ($Y_3Fe_5O_{12}$ [52, 53]), and bonding length modification ($SmB_6$ [54]). Examples of metallization without involving any structural changes are rare as far as we know. Examples of irreversible metallization were only found in some layered materials such as $MoS_2$ [55], $MoSe_2$ [56], α-$As_2Te_3$ [57], and $Sb_2S_3$ [58], where the irreversible metallization was attributed to permanent plastic deformation of interlayer spacing under nonhydrostatic pressure. Note that our FeSi sample is not a layered material. However, it was visibly flattened upon increasing pressure and remained so after it was removed from the DAC, indicating an irreversible plastic deformation.

There have been some reports of magnetic ordering in the surface state [23, 25, 59, 60] or nanowires [61] of FeSi. Considering FeSi as a TKI candidate, we proposed [23] the magnetically ordered CSS could result from the effect of Kondo breakdown on the FeSi surface [62]. From the decreasing area of the *MR* loop under pressure, we inferred the progressive suppression of the magnetic ordering and its final disappearance simultaneously with the metallization of the semiconducting bulk material. In terms of a Kondo model, this process could be visualized in terms of pressure-induced breaking of Kondo singlets consisting of Fe magnetic moments and antiferromagnetically bound electrons in FeSi. These liberated electrons become itinerant which leads to the metallic state under pressure. More magnetic measurements are required to ascertain whether metallic FeSi exhibits long-range magnetic ordering as in $SmB_6$ [11] and golden-phase SmS [10, 12].

**Concluding Remarks**

High pressure electrical resistance experiments revealed a semiconducting to metallic bulk transition in FeSi at $P_{cr}$ ~ 10 GPa, supported by the dramatic decrease of the energy gap and room temperature electrical resistance. Upon release of pressure, the metallic state persists to near ambient pressure. The origin of this behavior, related to either a structural change or an electronic transition, is currently under investigation. Concurrently with the pressure-induced metallization, the suppression of hysteretic $R(B)$ of the CSS within ± 0.5 T and the negative-to-positive sign change in $dR/dB$ up to 9 T, suggest the disappearance of the magnetically ordered CSS, and further imply there is a strong connection between the CSS and the semiconducting bulk state in FeSi at low pressures below $P_{cr}$.

**Acknowledgements**

Research at the University of California, San Diego was supported by the National Nuclear Security Administration (NNSA) under the Stewardship Science Academic Alliance Program through the US DOE under Grant DE-NA0004086 (measurements at high pressures) and the US Department of Energy (DOE) Basic Energy Sciences under Grant DE-FG02-04ER46105 (crystal growth and characterization). This work was sponsored in part by the UC San Diego Materials Research Science and Engineering Center (UCSD MRSEC), supported by the National Science Foundation (NSF) under Grant DMR-2011924. Research (back-to-ambient-pressure x-ray diffraction) at Michigan State University was supported by the U.S. DOE-BES under Contract No. DE-SC0023648. JL, SS, and JH were supported by NSF DMREF-2118718. Development of portable ruby fluorescence system was supported by NSF CAREER-1453752. The designer diamond anvil was provided by the University of Alabama Birmingham, through NSF Grant DMR-2310526.